\documentclass[11pt]{evolution}    

\usepackage{natbib}
\usepackage{color}

\usepackage{epsfig,amsmath,amsfonts,amssymb,cases,subfigure}

\newcommand{\be}{\begin{equation}}
\newcommand{\ee}{\end{equation}}
\newcommand{\bea}{\begin{eqnarray}}
\newcommand{\eea}{\end{eqnarray}}
\newcommand{\bml}{\begin{mathletters}}
\newcommand{\eml}{\end{mathletters}}

\newcommand{\tc}{\textcolor}

\renewcommand{\citep}[1]{(\citealt{#1})}

\raggedright

\begin{document}

\title{Adaptive walks and distribution of beneficial fitness effects}
\author{Sarada Seetharaman$^{1}$%
       \email{Sarada Seetharaman - saradas@jncasr.ac.in}%
      and
         Kavita Jain$^{2,*}$%
         \email{Kavita Jain - jain@jncasr.ac.in}}

\address{%
    \iid(1)Theoretical Sciences Unit,
Jawaharlal Nehru Centre for Advanced Scientific Research, Jakkur P.O.,
Bangalore 560064, India\\
    \iid(2) Theoretical Sciences Unit and  Evolutionary and
Organismal Biology Unit,
Jawaharlal Nehru Centre for Advanced Scientific Research, Jakkur P.O.,
Bangalore 560064, India
}%

\maketitle
Running Title : Adaptive walk 

\vspace{5mm}
{\bf Contact Information (for all authors)}

Sarada Seetharaman

{\bf postal address}: Theoretical Sciences Unit,
Jawaharlal Nehru Centre for Advanced Scientific Research, Jakkur P.O.,
Bangalore 560064, India

{\bf work telephone number}: +91-80-22082967

{\bf E-mail}: saradas@jncasr.ac.in

Kavita Jain 

{\bf postal address}: Theoretical Sciences Unit and  Evolutionary and
Organismal Biology Unit,
Jawaharlal Nehru Centre for Advanced Scientific Research, Jakkur P.O.,
Bangalore 560064, India

{\bf work telephone number}: +91-80-22082948

{\bf E-mail}: jain@jncasr.ac.in

\clearpage
\begin{abstract}
We study the adaptation dynamics of a maladapted 
asexual population on rugged fitness landscapes with many local fitness peaks.  
The distribution of beneficial
fitness effects is assumed to belong to one of the three extreme value
domains, viz. Weibull, Gumbel and Fr{\'e}chet. We work in the
strong selection-weak mutation regime in which beneficial
mutations fix sequentially, and the population performs an uphill walk
on the fitness landscape until a local fitness peak is reached. A
striking prediction of our analysis is that the fitness difference between  
successive steps follows a pattern of diminishing returns in the Weibull
domain and accelerating returns in the Fr{\'e}chet domain, 
as the initial fitness of the population is
increased. These trends are found to be robust with respect to 
fitness correlations. We believe that this result can be exploited in
experiments to determine the extreme value domain of the distribution
of beneficial fitness effects. 
Our work here differs significantly from the previous ones that assume
the selection coefficient to be small.  
On taking large effect mutations into account, we find that 
the length of the walk shows different qualitative 
trends from those derived using small selection coefficient approximation. 

\vspace{1cm}

{\bf KEY WORDS:}  adaptive walk, distribution of beneficial fitnesses,
extreme value theory
\end{abstract}
\clearpage


The problem of adaptive evolution is challenging  because advantageous
mutations, which are responsible for adaptation, are rare
\citep{Eyrewalker:2007}.  However as beneficial mutations 
  contribute substantially to the  
  fate of a population inspite of their rarity, and play a crucial 
  role in real life scenarios such as the anti-drug resistance developed
  by microorganisms \citep{Bull:2005b}, it is  
  important to know the size and frequency of these mutations. In fact, a  
  fundamental question in the study of adaptive dynamics is 
  whether adaptation happens via many mutations conferring small
  fitness advantage, or a few producing large fitness
 changes. Although initial theoretical works
 suggested that adaptation occurs mostly by mutations that provide
 small benefits \citep{Fisher:1930a,Orr:2003a}, it has been recently
 realised that large effect mutations are
 also possible \citep{Joyce:2008}. The basic idea governing the shape
 of the distribution of beneficial fitness effects (DBFE) is due to
 \citet{Gillespie:1983}, who argued that in the event of a small
 environmental change, as the wild type fitness is expected to remain
 high, the mutations conferring higher fitness than the wildtype will 
lie in the  
right tail of the fitness distribution. The statistical
properties of such extreme fitnesses are described by an {\it extreme
  value theory} which states that  
the extreme value distribution of independent random variables can be
of three types: Weibull which 
occurs when the fitnesses are right-truncated, Gumbel for
distributions decaying faster than a power law and 
  Fr{\'e}chet for distributions with algebraic tails
  \citep{Sornette:2000}. During the last decade, the DBFE has been
  measured in several experiments on microbes
\citep{Sanjuan:2004a,Rokyta:2005,Kassen:2006,Rokyta:2008,Maclean:2009,Bataillon:2011,Schenk:2012}  
and  all the three extreme value domains have now been observed.

In recent years, the dynamics of adaptation have been studied extensively
in experiments \citep{Elena:2003a} and 
several quantities such as the fitness rank of the mutant at the first
adaptive step \citep{Rokyta:2005}, the number of adaptive substitutions
\citep{Rokyta:2009,Schoustra:2009,Gifford:2011,Sousa:2012}, mean
fitness fixed during adaptation
\citep{Schoustra:2009,Gifford:2011} and its dependence on the initial
fitness \citep{Maclean:2010,Gifford:2011,Sousa:2012} have been
measured. However the relation of these properties of
  adaptation dynamics 
to the tail of the fitness distribution and hence DBFE is not
clear. The purpose of 
this article is to elucidate this connection by a detailed study of
certain adaptation properties.

We study the process of adaptation 
in the framework of an {\it adaptive walk model}  
\citep{Gillespie:1983,Gillespie:1991} which has been a
subject of many theoretical studies \citep{Orr:2002,Orr:2006a,Rokyta:2006,Joyce:2008, Kryazhimskiy:2009,Jain:2011d,
  Neidhart:2011, Jain:2011e,Filho:2012}. The  
  model is defined in the genotypic sequence space, and assumes strong
selection and weak mutation \citep{Gillespie:1983,Gillespie:1991}. These
conditions are met, for example, in natural populations of HIV-1 in
early infection 
\citep{daSilva:2012} and can also be designed in the laboratory
\citep{Sousa:2012}. In asexual populations under strong selection, a 
beneficial mutation gets 
fixed with a finite probability but the deleterious and neutral
mutations do not 
survive. Furthermore if the probability of mutation is small enough,
the population remains monomorphic at all times and its mutational
neighborhood is limited to single 
mutants. Such a population 
performs an adaptive walk on rugged fitness landscapes with many local
fitness peaks, in which fitness increases 
at each step until a local fitness optimum is reached since double and
higher order mutations are neglected.
Although experiments suggest
that the fitness landscapes are correlated
\citep{Carneiro:2010,Miller:2011,Szendro:2013}, most of the earlier
works on adaptive walks 
\citep{Rokyta:2006,Joyce:2008,Kryazhimskiy:2009,Neidhart:2011,Jain:2011e}
ignore correlations among fitnesses completely (however, see 
\citet{Orr:2006a,Jain:2011d,Filho:2012}).  
Here we model fitness 
correlations using a block model \citep{Perelson:1995} in which a sequence is assumed to be composed of independent partitions. 
Building upon a formalism introduced in \citet{Flyvbjerg:1992} and 
developed in \citet{Jain:2011d}, and using ideas from extreme value
theory \citep{Sornette:2000}, we study the evolution of fitness
and selection coefficient during the adaptive walk, and the number of
adaptive steps taken 
until the walk terminates.

We find that at the first few steps of the adaptation process, {\it fitness
difference} between successive adaptive substitutions displays a 
qualitatively different pattern in the three extreme value domains: it 
decreases in the Weibull domain, increases in the Fr{\'e}chet domain
and remains a constant in the Gumbel domain, as the initial fitness is
increased. This property is seen to hold for both uncorrelated and
correlated fitnesses. Since the fitness 
benefits conferred during the early adaptation stage are accessible in
experiments
\citep{Rokyta:2005,Schoustra:2009,Maclean:2010,Gifford:2011,Sousa:2012},
we believe that this result provides a simple way to determine the
extreme value domain of the DBFE.

We also find that the magnitude of fixed selective effects differs in
the three extreme value domains with small selection coefficients
occurring in the Weibull and Gumbel domains, and large ones in the 
Fr{\'e}chet domain. Previous studies
\citep{Orr:2002,Orr:2006a,Rokyta:2006,Joyce:2008,  
  Kryazhimskiy:2009,Jain:2011d, Neidhart:2011,
  Jain:2011e,Filho:2012} on adaptive walks  work with the assumption
that the selection coefficients are small. However large 
selection coefficients have been seen in 
experiments \citep{Bull:2000,Barrett:2006b} and so far,
very few theoretical investigations have taken large effect mutations
into account \citep{Heffernan:2002,Barrett:2006a}. 
Here we relax the assumption of small selective effects, and find that
large selection coefficients strongly affect the  
average number of adaptive
substitutions fixed during the adaptive walk. Our numerical
simulations show that the length of the adaptive walk is shortest in
the Gumbel domain. In contrast, within the small selection coefficient
approximation, it 
has been shown analytically that the walk length in the Fr{\'e}chet
domain is shorter than that in the Weibull and Gumbel domains   
\citep{Jain:2011d, Neidhart:2011,Jain:2011e}.


\section*{Models}

\subsection*{Block model of rugged fitness landscapes} 

We study adaptation on rugged fitness landscapes that are
characterised by many local fitness maxima using a \textit{block
  model} \citep{Perelson:1995} in which 
a sequence of length $L$ is split into $B$ blocks of equal length 
$L_B=L/B$. The partitioning of a sequence is motivated by the domain
structure of proteins \citep{Ponting:2002} and paired-unpaired regions
in RNA secondary structure \citep{Batey:1999}. In 
  proteins, the domains that 
  perform essential enzymatic functions are more likely to be stable
  and in RNA   
  secondary structure, the paired regions may have a lower free energy
  than the unpaired ones. 
In general, different blocks in a sequence may have different fitness
and a random variable chosen from a fitness 
distribution $p(f)$ may be assigned to each block. If interactions
between the blocks are neglected \citep{Ponting:2002}, the fitness of
the whole  
sequence can be written as the average of the block fitnesses
\citep{Perelson:1995}. 

For $B > 1$, two sequences with one or more common 
blocks have correlated fitness. For example, a sequence that is 
single mutation away from the wild type will have the same block
fitness in all but one block. 
As a result, the mutant fitness is close (or correlated) to that of
the parent sequence. The fitness correlations increase with 
the number of blocks in a sequence \citep{Perelson:1995,Das:2010}. For
$B=L$, we have the familiar additive fitness landscape in which fitnesses are
  completely correlated \citep{Kauffman:1993}, while for $B=1$, a 
completely uncorrelated fitness landscape is obtained
\citep{Kauffman:1993,Jain:2007a} in which even a single mutation can
result in a fitness completely different from that of the
parent sequence. Although 
such fitness landscapes are biologically unrealistic, they serve as a
useful starting point and as discussed later, the qualitative properties
of adaptation dynamics hold for both uncorrelated and correlated
fitnesses. Except for $B=L$, the fitness landscapes are rugged with
many local fitness optima that are sequences fitter
than all of their one mutant neighbours. 
For fixed $L$, since the number of local fitness peaks 
  decreases with increasing $B$ \citep{Perelson:1995}, fitness
  landscape gets smoother with increasing 
  correlations. Due to the presence of local fitness optima, these
  fitness landscapes exhibit sign epistasis which refers to the
  dependence of the beneficial or deleterious effect of a mutation on the 
  genetic background \citep{Weinreich:2005a,Poelwijk:2007,Jain:2011a}.

In order to specify the fitness distribution $p(f)$, one can exploit the fact
that adaptation occurs via rare beneficial mutations whose fitness
lies in the upper tail of the fitness distribution
\citep{Gillespie:1983,Gillespie:1991}. Then according to the extreme value
theory for independent random variables, the distribution of 
advantageous mutations can be one 
of the three types namely Weibull,
Gumbel and Fr{\'ec}het \citep{Sornette:2000}. It should be noted that the above classification of extreme value
domains for independent random variables is unlikely to hold for
strongly correlated fitnesses, but for weak correlations, one may still expect
it to work \citep{Clusel:2008,Jain:2011b}. Following
\citet{Joyce:2008}, we choose the block fitnesses from a generalised Pareto
distribution defined as  
\be
p(f)= (1+\kappa f)^{-\frac{1+\kappa}{\kappa}}
\label{gpd}
\ee
where the exponent $\kappa$ can take any real value. The
fitness is unbounded for $\kappa\geq0$ and for $\kappa<0$, it has an
upper bound $u$ at $-1/\kappa$. A nice feature of distribution
(\ref{gpd}) is that all the three extreme value domains can be
accessed by tuning a 
single parameter $\kappa$ with  $\kappa<0$, $\rightarrow 0$ and $>0$
leading to Weibull, Gumbel and Fr{\'e}chet distributions
respectively. 

A result from extreme value theory, which is 
relevant in the later 
discussion, states that  the typical
value $f$ of the $m$th best fitness amongst $L$ independent fitnesses can be determined by equating the
rank $m$ to the average number of fitnesses higher than $f$ which is given by
$L \int_{f}^u dg~p(g)$ \citep{Sornette:2000}. For the fitness
distribution (\ref{gpd}), this gives 
\be
L (1+\kappa f)^{-1/\kappa}=m 
\label{rank}
\ee
Setting $m=1$ in the above equation, we see that the typical
local peak fitness ${\tilde f}_B$  of a sequence of length $L$
partitioned into $B$ blocks is given by
\be
{\tilde f}_B=\dfrac{L_B^\kappa-1}{\kappa}
\label{largestB}
\ee
since it is the average of $B$ random variables, each of which is
the best of $L_B$ random variables. In the following discussion, we
will omit the subscript $1$ when referring to quantities for
uncorrelated fitnesses. For later reference, we also note that the mean 
of the fitness distribution 
(\ref{gpd}) is infinite for $\kappa \geq 1$ and the variance for
$\kappa \geq 1/2$. 


\subsection*{Adaptive walk model}

We consider a population of self-replicating binary 
sequences, each of length $L$ evolving in the \textit{strong selection-weak
  mutation} regime \citep{Gillespie:1983,Gillespie:1991}. The 
population is assumed to have a fixed size $N$, and mutation 
occurs with a probability $\mu$ per site per generation. 
In the weak mutation regime, 
the average number of single-mutants of a particular type 
produced per generation is smaller than one ($N \mu \ll 1$). Since the
time to generate sequences that 
are two mutations away is of the order $\mu^{-2}$
\citep{Orr:2002,Iwasa:2004}, which is much 
larger than that for a single mutation, we work on time scales over
which double and higher order mutants can be ignored and consider only
$L$ single mutants of a sequence. If selection is strong relative
to random genetic drift, while the neutral
and deleterious mutations get lost, a  
beneficial mutation with selection coefficient $s$ is fixed 
with a probability $\pi(s) \approx 1-e^{-2 s}$ 
\citep{Kimura:1962,Orr:2002}. Thus strongly 
selected mutants are more likely to get fixed than the weaker ones and
mutants with very large selective effects are almost certain to be
fixed. 

Under these conditions, in a maladapted population, although a single-mutant
with selection coefficient $s$ arises on an average every $(N \mu)^{-1}$
generations, the waiting time to its fixation is $(N \mu
\pi(s))^{-1}$ generations. 
If the initial fitness is small, several beneficial alleles each with
a different  
selective effect are possible, and Gillespie showed the probability
that the one with selection 
coefficient $s$ will sweep through the population is 
proportional to $\pi(s)$ \citep{Gillespie:1991}. Once a beneficial
mutant is fixed in the population, the new 
wild type produces
a {\it novel} neighborhood of mutants that are single mutation away
from it. Again one of the beneficial mutants sweeps through the
population and replaces the current wild type. This substitution
process goes on until the population 
encounters a local fitness peak, 
as double and higher order mutants are ignored.   

If the population is fixed at a
sequence with fitness $h$, a mutant with fitness $f > h$ substitutes
it with a probability proportional to the fixation probability
$\pi(s)$ where $s=(f-h)/h$. For long sequences, the normalised transition
probability is given by \citep{Jain:2011d} 
\be
T(f \leftarrow h) = \frac{(1-e^{- \frac{2 (f-h)}{h}}) ~p(f)}{\int_h^u
  dg~(1-e^{-\frac{2 (g-h)}{h}})~p(g)} ~,~f > h
\label{Tp}
\ee
where $u$ is the upper bound of the fitness distribution
$p(f)$. It is important to note that unlike the previous
  works \citep{Orr:2002,Orr:2006a,Rokyta:2006,Joyce:2008,   
  Kryazhimskiy:2009,Jain:2011d, Neidhart:2011,Jain:2011e,Filho:2012}
  that assume selection coefficient to be small, here we employ the
  full expression (\ref{Tp}).

In computer simulations of the dynamics of the adaptive walk, we started
with a sequence 
of length $L$ and initial fitness $f_0$, and considered uncorrelated
($B=1$) and weakly correlated fitnesses ($B=2$). In 
the former case, the initial fitness of the sequence is fixed and in
the latter case, two random variables are 
generated independently from (\ref{gpd}) and  they are accepted
as block fitnesses if their sum is $2 f_0 \pm \delta$ where $\delta
\sim 0.01 f_0$. At each step of the adaptive walk, we generate 
$L$ new fitnesses that are chosen from (\ref{gpd}) and one of them is
  chosen to be fixed according to the 
  transition probability (\ref{Tp}). While in the case of uncorrelated
  fitnesses, the fitness of the whole sequence changes at each step,
  when $B=2$ the fitness of only one of the blocks is changed. The
fitnesses sampled during the walk are 
not stored as for large $L$, the 
number of one mutant neighbors probed in previous steps can be
ignored in comparison to $L$
\citep{Orr:2002,Flyvbjerg:1992,Seetharaman:2011}. In our simulations, the
fitness and selection coefficient of each step are 
averaged over only those walks that proceed until that step. In all
the simulations on uncorrelated fitness landscapes, the data
were averaged over $10^6$ independent realisations of the fitness
landscape and $10^5$ for the correlated ones. 


\section*{Results}


\subsection*{Evolution of fitness fixed} 

\begin{figure}[ht]
\centering
\includegraphics[width=0.9
  \linewidth,angle=270]{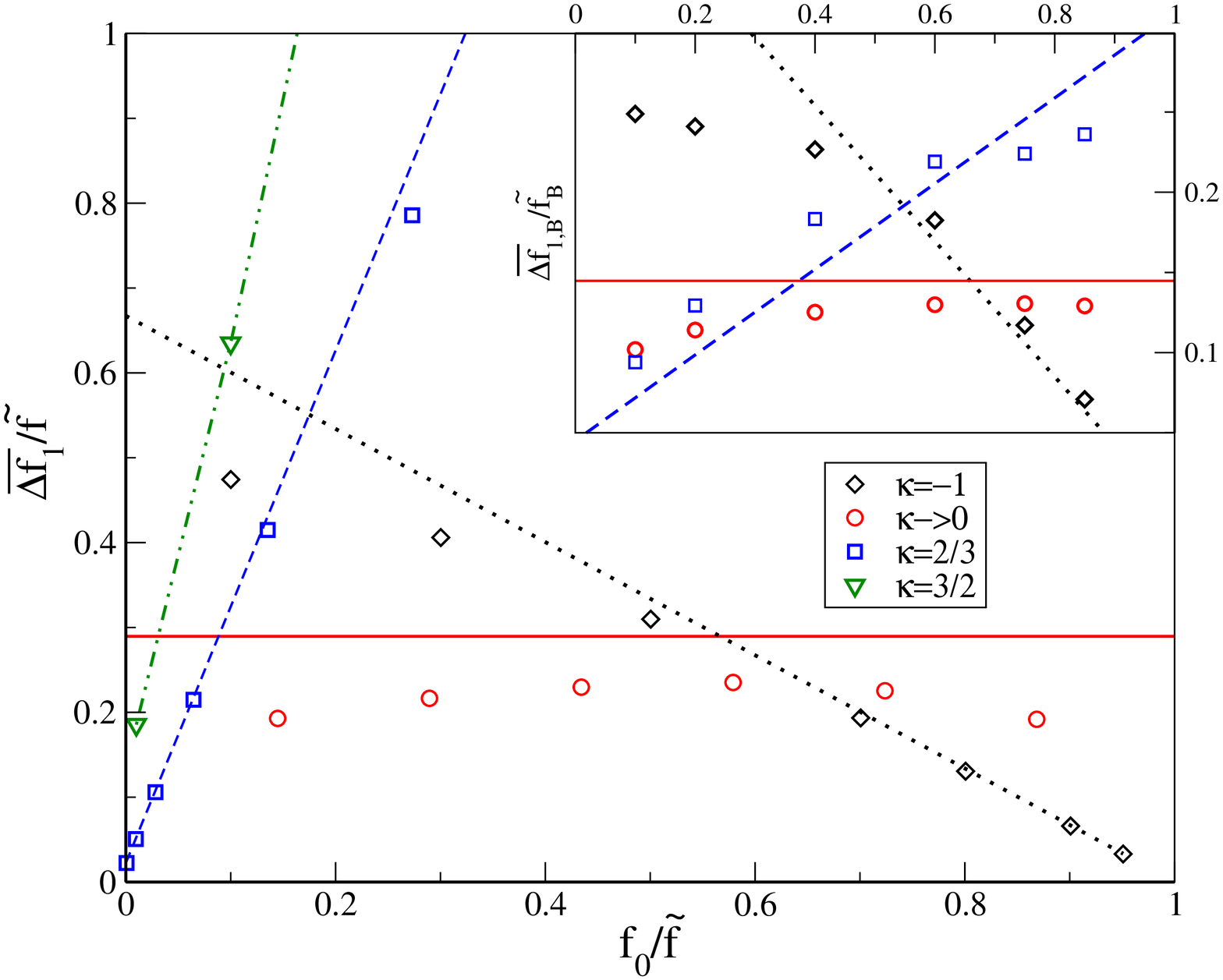}
\caption{The plot shows (scaled) average fitness difference at the
  first step as a function of initial fitness for various $\kappa$ on
  uncorrelated (main) and correlated fitness landscapes with $B=2$
  (inset). In both the plots, $L_B=1000$ which corresponds to $L=1000$
  and $2000$ for uncorrelated and correlated fitnesses respectively. The points
give the simulation data and the line connecting the data points are
obtained from 
(\ref{rateinfL}) and (\ref{corrfitdiff}) for uncorrelated and
correlated fitnesses respectively for $\kappa < 1$. The data points
for $\kappa=3/2$ are scaled down by $10^2$ for clarity and the line
connecting the data is guide to the eye.}
\label{Fig1}
\end{figure}

\begin{figure}
\centering
\includegraphics[width=0.9 \linewidth,angle=270]{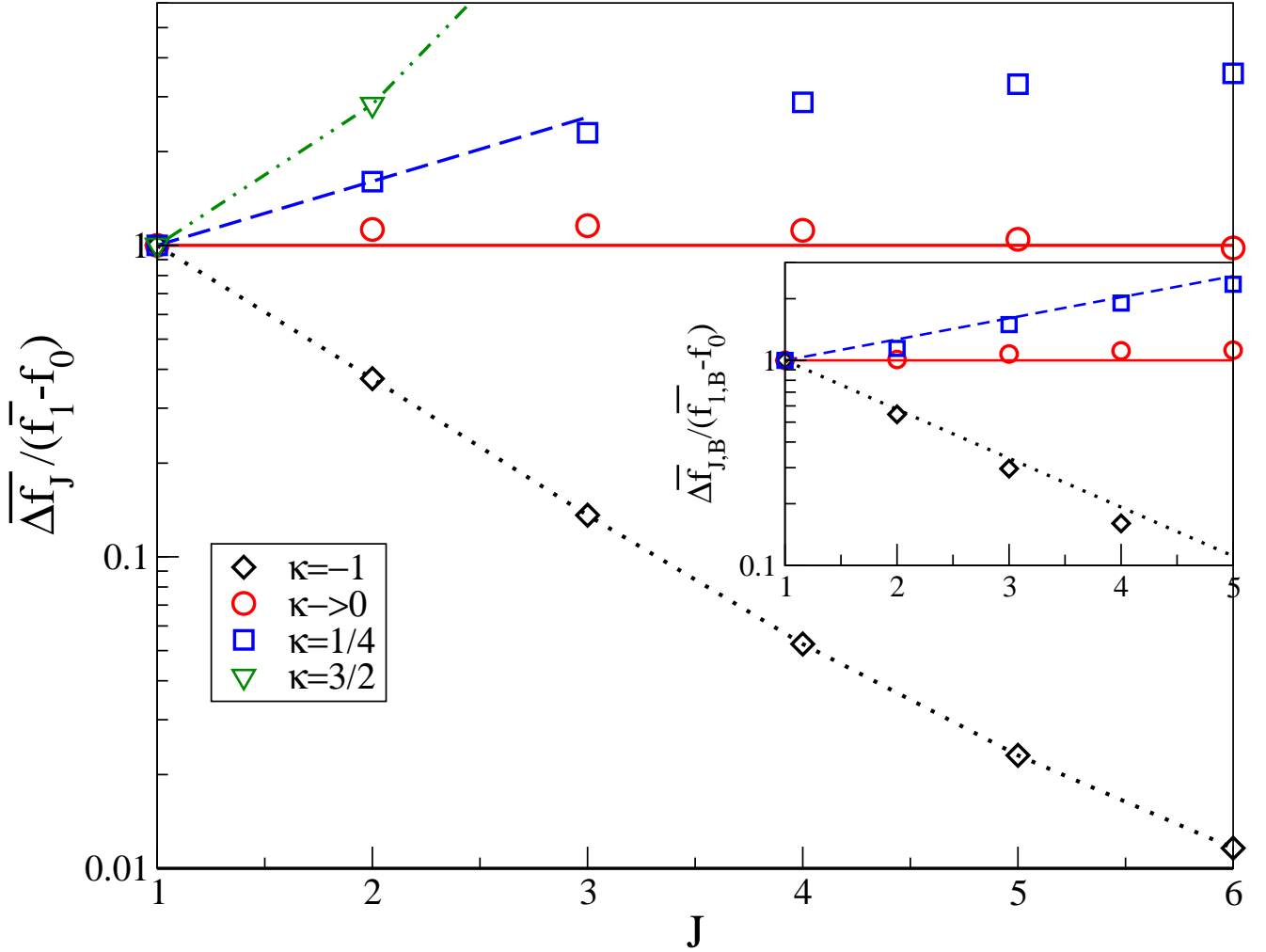}
\caption{The plot shows (scaled) average fitness difference between
  successive steps as a function of the number of adaptive
  substitutions for various $\kappa$ on uncorrelated (main) and
  correlated fitness landscapes with $B=2$ (inset). Taking
  $f_0=0.63,1,1.14$ and $2.32$ for $\kappa=-1,0,1/4$ and $3/2$
  respectively, the simulation data are shown as points for
  $L_B=1000$ which corresponds to $L=1000$ and $2000$ for independent
  and correlated fitnesses respectively. The line connecting the data
  points for $\kappa=3/2$ is guide to the eye, while the others are
  obtained from (\ref{rateinfL}) and (\ref{corrfitdiff}) for
  uncorrelated and correlated fitnesses respectively.} 
\label{Fig2}
\end{figure}

\begin{figure}
 \centering
\includegraphics[width=0.9 \linewidth,angle=270]{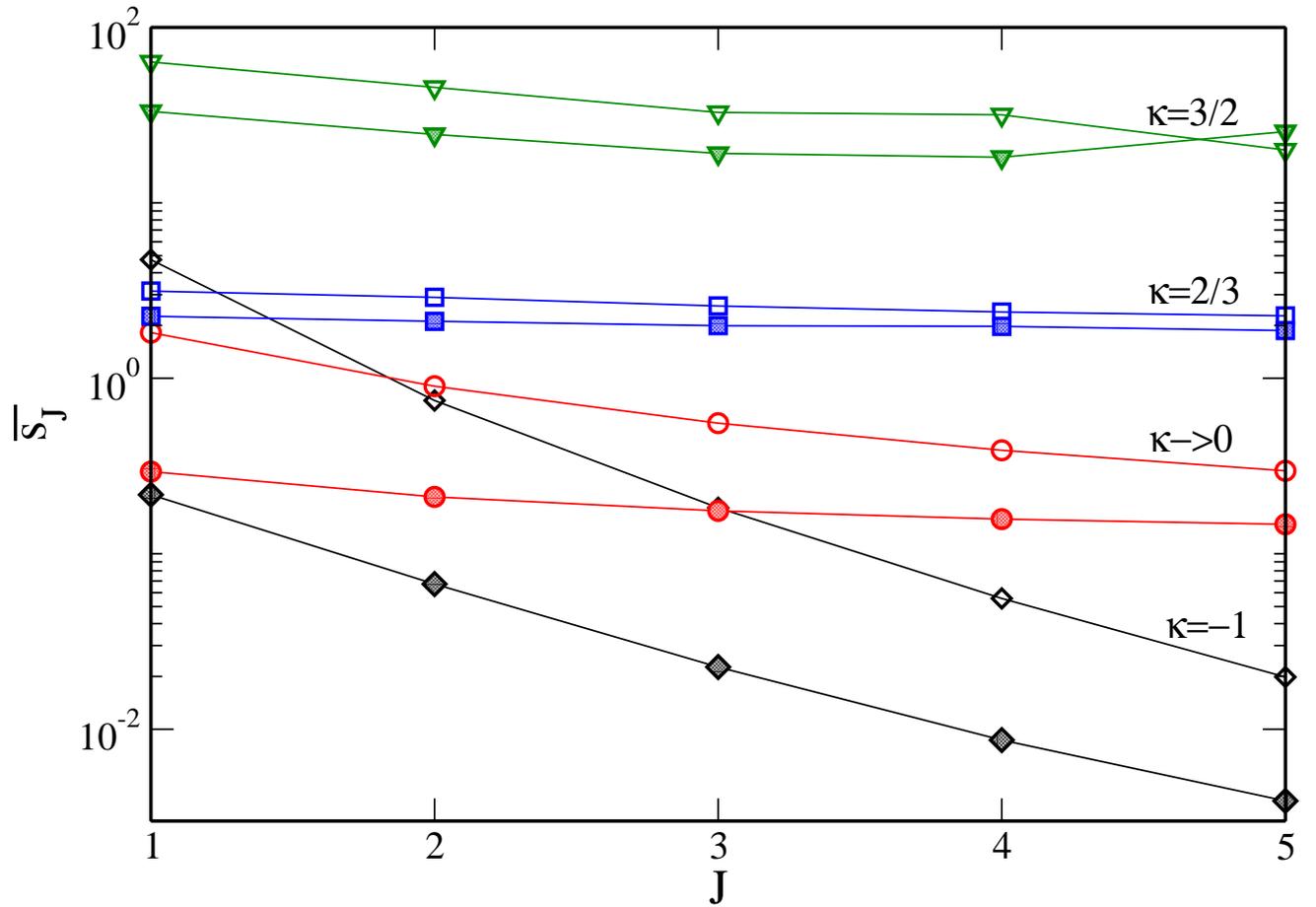}
\caption{The plot shows the average selection coefficient fixed during the
  course of the walk on uncorrelated fitness landscapes for various
  $\kappa$ and $L=1000$. The open and shaded symbols are respectively for
  $f_0=0.1 \tilde {f} \text{ and } 0.75 \tilde {f}$ where $\tilde
  {f}$ is the average fitness of a local fitness peak given by
  (\ref{largestB}). The points are the simulation data, while the lines
  are guide to the eye. The data for $\kappa=3/2$ is scaled down
  by a factor $10$ for clarity.}
\label{Fig3}
\end{figure}

 While the fitness fixed during adaptation increases in
  all the three extreme value domains \citep{Jain:2011d}, the average 
  difference $\overline{\Delta f_J}$ between fitnesses fixed at step
  $J-1$ and $J$ exhibits interesting trends that 
  can be exploited to distinguish between them, refer
  Figs.~\ref{Fig1} and \ref{Fig2}. We find that 
  $\overline{\Delta f_J}$ decreases during the walk in the Weibull
  domain and increases in the Fr{\'e}chet domain. 
A similar behavior is seen at a fixed step in the walk when initial
fitness is varied. A heuristic understanding of the latter result can be
obtained for uncorrelated fitnesses using a simple
back-of-the-envelope calculation of the 
average fitness ${\bar f}_1$ at the first step, which is given by $\int_{f_0}^u
df~f~T(f \leftarrow f_0)$. We first note that if the fitness
distribution decays slowly, fitnesses much larger than initial
fitness can occur with appreciable frequency and thus the selection
coefficients can be large. On the other hand, for bounded
distributions, the selection coefficient can be at most $u/f_0-1$ which is below
unity for $f_0 > u/2$. Indeed as Fig.~\ref{Fig3} shows, the selection
coefficients fixed are large (small) for positive (negative)
$\kappa$. As a result, the fixation probability $\pi(s)$ can be
approximated by unity in the 
Fr{\'e}chet domain, while $\pi(s) \approx 2 s$ in the Weibull domain. A quick
calculation gives ${\bar f}_1 \sim f_0/(1-2 \kappa)~,~\kappa < 0$ which is
linear in $f_0$ with a slope below unity. On the
other hand, in the Fr{\'e}chet domain, a transition occurs in the
behavior of the fitness fixed at $\kappa=1$ where the mean of the
distribution $p(f)$ becomes infinite. We find that the average fitness
is infinite for $\kappa \geq 1$ but for $0 < \kappa < 1$, 
the fitness ${\bar 
  f}_1 \sim f_0/(1-\kappa)$ which also increases with $f_0$ but with a
slope above unity. The key
point that emerges from these simple calculations (and detailed ones
in Supporting information) is that the average fitness at the first step is of the
form $a f_0+b$ where the slope $a$ is above (below) one for positive
(negative) $\kappa$. The result for the fitness difference claimed
above then immediately follows.

 To understand the behavior at higher steps in the
  adaptive walk, more work is
  required and the detailed derivations are given in Supporting information. 
For infinitely long sequences, on using the results in Supporting information, we have 
\begin{subnumcases}
{\label{rateinfL} \overline{\Delta f_J}=}
a_-^{J-1} ~((a_--1) f_0+b_-)  ~,~ \kappa < 0 
\label{rateinfLW}\\
2 ~,~ \kappa \to 0 \label{rateinfLG} \\
a_+^{J-1} ~((a_+-1) f_0+b_+) ~,~ 0 < \kappa <  1 \label{rateinfLF}
\end{subnumcases}
where $a_- < 1, a_+ > 1$. For fixed  
initial fitness, the above equation shows that for $\kappa < 0$, the
fitness benefit decreases exponentially as the walk proceeds
({\it diminishing returns}) while for $\kappa \to 0$, the
fitness gain is same ({\it constant returns}) and 
for $0 < \kappa < 1$, it increases
exponentially fast with each step conferring higher benefit than
the previous one ({\it accelerating returns}). Similar qualitative
trends are seen when the initial fitness  is varied: the fitness
increment decreases (increases) linearly with $f_0$ for negative
(positive) $\kappa$. 
In Figs.~\ref{Fig1} and \ref{Fig2}, the
simulation results and the above theoretical prediction
(\ref{rateinfL}) for infinitely long sequence are compared and we see
a good agreement when the initial
fitness is sufficiently large but local fitness maximum is far
away. The latter condition is satisfied when the number of adaptive 
substitutions and the initial fitness are smaller than the average
length of the walk and the average fitness of a local maximum
respectively. The results of our numerical simulations in
Figs.~\ref{Fig1} and \ref{Fig2} also show that the fitness 
difference between successive steps increases with both  $f_0$ and $J$
for $\kappa \geq 1$ as well. 

It is instructive to compare the expression (\ref{rateinfL})
obtained using (\ref{Tp}) with the one 
that assumes small selection coefficient. In the {\it small 
  selection coefficient approximation}, the transition 
probability (\ref{Tp}) reduces to 
\be
T(f \leftarrow h)=\frac{(f-h) p(f)}{\int_h^u dg~ (g-h)~p(g)}~,~f > h
\label{linapprox}
\ee 
A straightforward calculation carried along the lines described in Supporting information  
shows that the fitness difference is given by (\ref{rateinfLW}) for all
$\kappa < 1/2$ or more explicitly, 
\bea
\widetilde{\overline{\Delta f_J}}= 2 (1+\kappa f_0) (1-2
\kappa)^{-J}~,~\kappa < 1/2
\label{rateinfL2}
\eea
where the `$\sim$' denotes the result obtained within small
selection coefficient approximation. The condition $\kappa < 1/2$  in
(\ref{rateinfL2}) arises due to the infinite variance of the fitness
distribution 
(\ref{gpd}). An expression similar to (\ref{rateinfL2}) for fitness
effects has been obtained by 
\citet{Joyce:2008} but its consequences were not discussed. The above
result has also been obtained for the special case of exponentially
distributed fitnesses and zero initial fitness by 
\citet{Kryazhimskiy:2009}. From (\ref{rateinfL}) and 
(\ref{rateinfL2}), we first note that in all the three extreme value 
domains, fitness difference displays the same qualitative trend,  
irrespective of whether the correct asymptotic behavior of transition
probability is taken into account. However the result 
(\ref{rateinfL2}) matches  with (\ref{rateinfL}) 
in the Weibull and Gumbel 
domains but not in the Fr{\'e}chet domain. 
This is because the selection coefficient, shown in 
Fig.~\ref{Fig3} for two initial fitnesses, decreases with $f_0$ for
$\kappa \leq 0$ and
at sufficiently large $f_0$, selective effects can be assumed to be small.  
But for $\kappa > 0$, the selection coefficient remains
high even for large initial fitnesses and therefore we do not expect
the small selection coefficient approximation to work here. The
behavior of the selection coefficient 
can be immediately obtained at the first step in the walk using (\ref{rateinfLW})-(\ref{rateinfLF})
since ${\bar s}_1=\overline{\Delta f_1}/f_0$ and we find that ${\bar s}_1$
decays to zero with increasing $f_0$ 
for $\kappa \leq 0$, but to a finite constant $a_+-1$ for $0 < \kappa <
1$. On comparing (\ref{rateinfLF}) and
(\ref{rateinfL2}), we find that the value of the exponent $\kappa$ at
which a transition occurs in the behavior of the fitness fixed is
different. Moreover the growth rate 
$a_+$ (given in Supporting information), which takes values in the range $1.1-27.5$
as $\kappa$ is increased 
from $0.05$ to $0.95$, is smaller than the corresponding rate $(1-2
\kappa)^{-1}$ in (\ref{rateinfL2}) because the transition probability
(\ref{Tp}) decays faster than (\ref{linapprox}) for large fitnesses.

\tc{black}{When a sequence is partitioned into $B$ blocks, the fitness
  of the sequence at any step is determined by the {\it joint} 
distribution of the fitness of
the block that acquired one beneficial mutation and the fitnesses of
the rest of the blocks at the preceding step. But as it is difficult
to work with this distribution, here we use the approximation that the
joint distribution can be factorised over the blocks. In other words,
we assume that the blocks evolve independently which is a reasonable
approximation for weakly correlated fitnesses. Since $J$ substitutions in 
a sequence partitioned in $B$ blocks can be obtained if each block
acquires $J/B$ mutations, it immediately follows that 
\be
\overline{\Delta f_{J,B}}\approx {\bar f}_{J/B}-{\bar f}_{(J/B)-1} ~,~J > 0
\label{corrfitdiff}
\ee
where ${\bar f}_J$ is the average sequence fitness at the $J$th step
on uncorrelated fitness landscapes. 
Using the results of Supporting information in the above equation, we find 
that the trend of fitness difference for correlated fitnesses 
is the same as that in the uncorrelated case. This result is
consistent with the simulation data shown in  
the inset of Figs.~\ref{Fig1} and \ref{Fig2} where the fitness
difference increases and decreases for $\kappa>0$ and $< 0$
respectively. 
The selection coefficient decreases with increasing correlations in
all extreme value domains. Our numerical data, for a parameter set in
which the same   
initial and final fitness is chosen for uncorrelated and correlated
fitnesses, shows that the selection coefficient at the first step
reduced from $1.4$ to $0.8$ for $\kappa=-1$, $0.7$ to $0.4$ for
$\kappa \to 0$ and $2.8$ to $0.4$ for $\kappa=2/3$, as the block number 
increased from one to two.}

\subsection*{Average length of the walk}

\begin{figure}[ht]
\centering
\includegraphics[width=0.9
  \linewidth,angle=270]{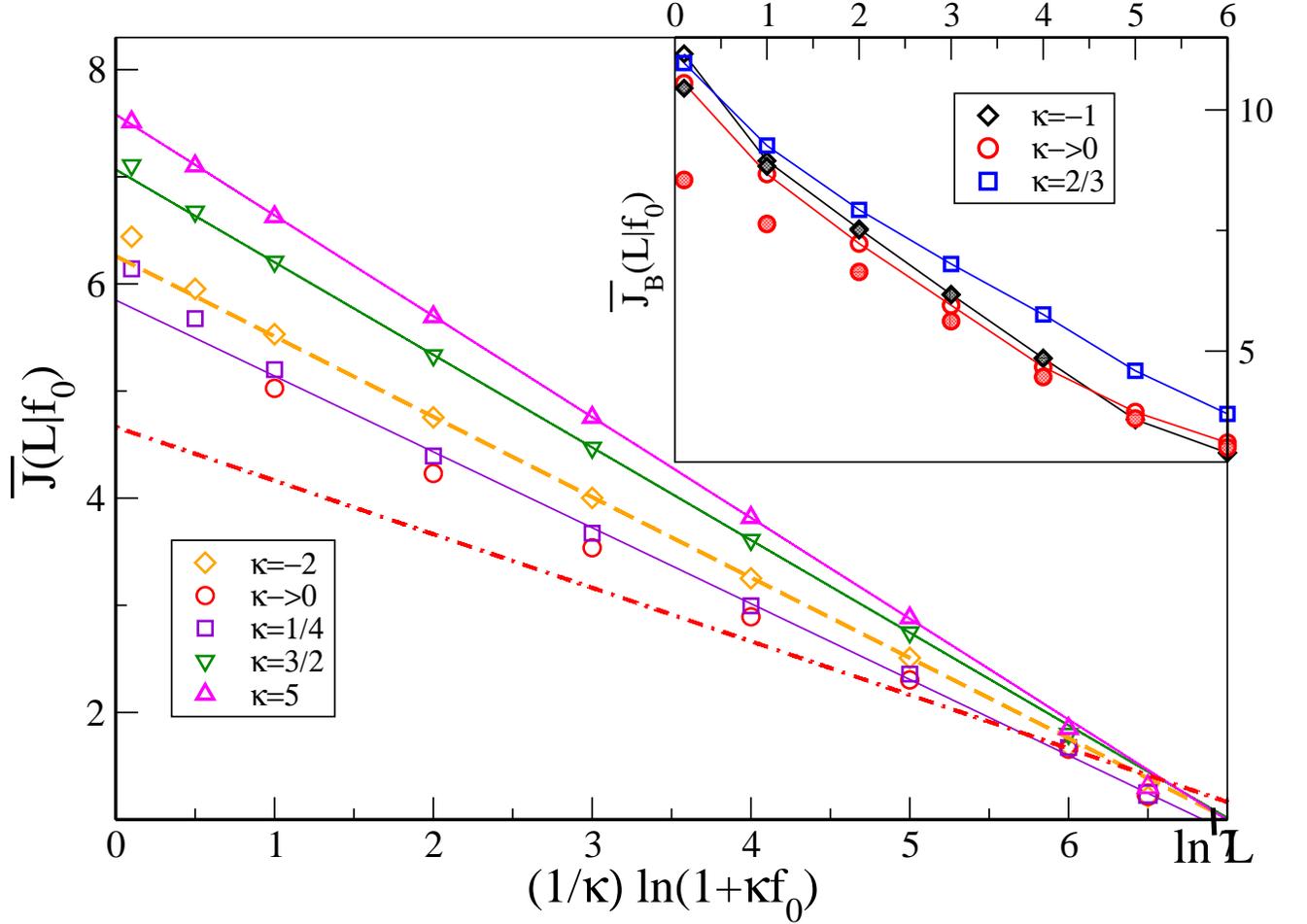}
\caption{The plot shows  the variation of the average
  walk length with initial fitness for various $\kappa$ on uncorrelated (main) and correlated fitness landscapes with $B=2$ (inset). In the main
  plot, the broken lines show the result (\ref{wlfit})
with the constants ${\tilde c}=1.08$ and $1.21$ for $\kappa=-2$ and
$\rightarrow 0$
  respectively, while the solid
  lines are the best fit to (\ref{rank_wl}) with $\beta \approx 0.71,
  0.86, 0.94$ for $\kappa=1/4, 3/2$ and $5$ respectively.
In the inset, the open symbols give
  the simulation data points of the average walk length obtained using
  the transition probability (\ref{Tp}) while the shaded ones are
  those obtained using the transition probability
  (\ref{linapprox}) in the small selection coefficient
  approximation. In all the simulations, the sequence length $L=1000$.}    
\label{Fig4}
\end{figure}

The number of adaptive substitutions that occur as the population
moves from the initial fitness to a local fitness peak is termed the
\textit{walk length}. For an infinitely long sequence, the walk goes
on forever for all $\kappa$ \citep{Jain:2011d} but for finite $L$, the
walk terminates at a local fitness peak and the walk length is
expected to increase with the sequence length.  Here we are unable to
analytically calculate the average walk length 
and present our numerical results in Fig. \ref{Fig4}. For uncorrelated
fitnesses, we find that in all the extreme value domains, the average
walk length decreases with increasing $f_0$ due to decreasing
availability of beneficial mutations at higher initial fitnesses. The
simulation results also indicate that the average walk length ${\bar
  J}(L|f_0)$ has a logarithmic dependence on the rank $m_0$ of the
initial fitness which, by virtue of (\ref{rank}), is given by
$m_0=L(1+\kappa f_0)^{-1/\kappa}$. Thus we can write 
\be
{\bar J}(L|f_0)=\beta \ln m_0+c
\label{rank_wl}
\ee
where  $\beta$ and $c$ depend on the exponent $\kappa$ and the block
number $B$. Interestingly, the prefactor $\beta$ has a
\textit{nonmonotonic} dependence on the exponent $\kappa$:  with
increasing $\kappa$, it decreases in the Weibull domain and increases
in the Fr{\'e}chet domain with a minimum occurring in the Gumbel 
domain.  
As shown in the inset of Fig.~\ref{Fig4}, on correlated fitness
landscapes, the adaptive walks are longer than those on uncorrelated
ones since the number of local fitness peaks decrease with increasing
correlations \citep{Orr:2006a}. Furthermore the average walk length ${\bar
  J_B}(L|f_0)$ seems roughly linear in $\ln m_0$ in all the three
extreme value domains with a slope that depends nonmonotonically on
exponent $\kappa$.  

In the previous section, we saw that the behaviour of fitness fixed
can be understood using the small selection coefficient approximation
in the Weibull and Gumbel domains. Below we will compare our results in
Fig. \ref{Fig4} with those obtained assuming that the selective
effects are small.  
Using (\ref{linapprox}), analytical expressions for average walk
length have been obtained for both uncorrelated and correlated
fitnesses \citep{Jain:2011d,Neidhart:2011,Jain:2011e,Seetharaman:2013}
and it has been shown that a transition occurs in the behaviour of the
walk length at $\kappa=1$. For  $\kappa<1$ where the mean of the
fitness distribution (\ref{gpd}) is finite, the average walk length calculated 
using the transition probability (\ref{linapprox}) is found to be  
\be
{\tilde{\bar{J}}}(L|f_0)  = \tilde{\beta} \ln m_0 +\tilde{c}~,~\kappa<1
\label{wlfit}
\ee
where  $\tilde c$ is a constant, and on uncorrelated fitness
landscapes \citep{Jain:2011d,Neidhart:2011,Jain:2011e},
${\tilde\beta}$ is given by 
\be
{\tilde\beta}=\dfrac{1-\kappa}{2-\kappa} ~,~\kappa<1
\label{beta}
\ee
An intuitive understanding of the logarithmic dependence of the walk
length in the domain of $\kappa$ where its variance is finite can be
obtained by equating the fitness fixed at the final step ${\tilde
  {\bar J}}$ to the average fitness
(\ref{largestB}) of a local fitness maximum
\citep{Flyvbjerg:1992,Jain:2011d}. Note that ${\tilde\beta}$ in
(\ref{beta}) decreases monotonically with the exponent $\kappa$.  For
$\kappa>1$ where the mean of the fitness distribution (\ref{gpd}) is
undefined, the walk length is found to be independent of the initial
fitness rank.  Our analytical calculations for the average walk length on
correlated fitness landscapes \citep{Seetharaman:2013} show that the
length of the adaptive walk increases with increasing block number $B$
and there is a monotonic decrease in $\tilde \beta$ with increasing
$\kappa$.

We now exploit the results summarised above to understand the walk
length behaviour when the transition probability is given by
(\ref{Tp}). Figure \ref{Fig4} shows that in the Weibull domain, for
uncorrelated fitnesses, the simulation data and the
expression (\ref{wlfit}) are in good agreement and for correlated
fitnesses, when the numerical data obtained using (\ref{Tp}) is
plotted with the ones using (\ref{linapprox}), the two data sets
coincide for a wide range of initial fitness. Similar plots in the
Gumbel domain for uncorrelated and correlated fitnesses show that the
small selection coefficient approximation does not work as well as in
the Weibull domain.  
In the Fr{\'e}chet domain, the results (\ref{wlfit}) and (\ref{beta})
obtained by neglecting the large effect mutations predict decreasing
walk length with increasing $\kappa$, a trend opposite to that seen in
Fig.~\ref{Fig4}.  That the walk length ${\bar J}$ should be longer
than ${\tilde {\bar J}}$ is expected - as the step length
(\ref{rateinfLF}) is smaller than that given by (\ref{rateinfL2}),
more adaptive steps to a local fitness peak can be taken in
the former case. To understand the trend of coefficient $\beta$, it is
useful to consider the limits $\kappa \to \pm \infty$. It has been
shown that the limit $\kappa \to -\infty$ corresponds to a random
adaptive walk in which transition to any beneficial mutation occurs
with the same probability  \citep{Joyce:2008} and the walk length is
given by (\ref{wlfit}) with ${\tilde \beta}=1$
\citep{Flyvbjerg:1992}. The opposite limit $\kappa \to \infty$
corresponds to a greedy adaptive walk
\citep{Orr:2003b,Campos:2005} in which the fittest mutant
is chosen with probability one if selection coefficient is assumed to
be small, but a random adaptive walk when large selective effects are
taken into consideration \citep{Joyce:2008}. Thus we arrive at the
conclusion that in the two limiting cases, if the selection
coefficient is allowed to be large, the prefactor $\beta$ in
(\ref{rank_wl}) must be one. As the prefactor $\beta$ decreases with
increasing $\kappa$ due to (\ref{beta}) in the Weibull domain, it must
increase in the Fr{\'e}chet domain in order to satisfy the $\kappa \to
\infty$ limit. We also mention that the transition in 
the fitness fixed at $\kappa=1$ does not seem to affect the walk length.

\section*{Discussion}

In this article, we investigated how the statistical properties of the adaptive
walk relate to the tail behavior of the fitness distribution. 
The sign of the exponent $\kappa$ in the fitness 
distribution (\ref{gpd}) determines the nature of the DBFE which can
be of three types 
namely Weibull, Gumbel and Fr{\'e}chet. It is important to
  note that this 
classification of the extreme value domains is applicable only if the
fitnesses are completely uncorrelated or at most, weakly correlated
\citep{Clusel:2008}. For strongly correlated fitnesses, a
classification of extreme value domains on the basis of the behavior
of the tail of the fitness distribution is not available and it is not
clear if it even 
exists. For these reasons, here we studied the adaptive walk
properties on weakly correlated fitness landscapes \citep{Perelson:1995}.

The exponent $\kappa$ in (\ref{gpd}) has been measured in experiments and
interestingly, all the 
three extreme value domains for uncorrelated fitnesses have been
seen. Although many early  
studies supported the Gumbel domain
\citep{Sanjuan:2004a,Rokyta:2005,Kassen:2006,Maclean:2010}, 
recently Weibull \citep{Rokyta:2008,Bataillon:2011} and
Fr{\'e}chet domain \citep{Schenk:2012} have also been
documented. It has been suggested that as the beneficial mutations are
  expensive due to pleiotropic constraints, fat-tailed fitness
  distributions whose extreme value statistics lies in the
  Fr{\'e}chet domain can occur if such constraints are limited, and
  bounded 
  ones that lie in the Weibull domain for severe constraints
  \citep{Schenk:2012}.  
Experiments suggest that the fitness landscapes are correlated
\citep{Carneiro:2010,Miller:2011,Szendro:2013} but we know little 
about the fitness correlations quantitatively.  Sign epistasis which
is a characteristic feature of rugged fitness landscapes with many
local fitness maxima has also been
documented in several recent experiments
\citep{Poelwijk:2007,Franke:2011,Kvitek:2011,Lalic:2012}.

\subsection*{Comparison to previous works}

Our theoretical analysis here differs in an important way from the previous
studies on adaptive walks \citep{Orr:2002,Orr:2006a,Rokyta:2006,Joyce:2008,  
  Kryazhimskiy:2009,Jain:2011d, Neidhart:2011,
  Jain:2011e,Filho:2012} that assume selective effects to be small and
therefore work with (\ref{linapprox}) in which the transition probability is
linear in the selection coefficient. Here 
instead we work with (\ref{Tp}) which is a nonlinear function of the 
selection coefficient. Our numerical data in Fig.~\ref{Fig3} on
selection coefficient fixed shows that large selection coefficients
can arise in any extreme value   
domain when the initial fitness is small, as is the case in adaptation
experiments in stressful environment
\citep{Bull:2000,Barrett:2006b}, or in a moderately fit population if
the underlying fitness distribution is 
slowly decaying as is the case in the Fr{\'e}chet domain. Here we 
focus on the latter situation and therefore our formulae 
hold for moderately high initial fitnesses 
that are far from a local fitness optimum. 

Our main conclusion is that small selection 
coefficient approximation can be safely employed in the Weibull
domain, but the analysis in the Gumbel and especially Fr{\'e}chet domain
requires that large effect mutations are taken into
account. We find that regardless of whether we assume mutations to have small or
large effect, a transition in the behavior of the fitness fixed occurs at a
certain value of exponent $\kappa$ below which the fitness fixed
during the initial steps in the walk does not depend on the average
fitness of a local peak, and above which it does. The transition point
is given by $\kappa=1$ where mean of the 
fitness distribution becomes infinite if transition probability
(\ref{Tp}) is used but by $\kappa=1/2$ 
which corresponds to an infinite variance, if selection coefficient is
assumed to be small.  
More striking difference is observed in 
the length of the adaptive walk. For infinitely 
long sequences, the adaptive walk lasts forever for all $\kappa$
\citep{Jain:2011d} and as Fig.~\ref{Fig4} 
shows, the least number of adaptive substitutions occur
in the Gumbel domain. 
But assuming that the selective effects are small, a
transition is known to occur at $\kappa=1$ below which the walk length  
decreases as the exponent $\kappa$ is increased from Weibull to
Fr{\'e}chet domain \citep{Jain:2011d,Neidhart:2011,Jain:2011e}.

\subsection*{Evolution of fitness and selection coefficient}

Although the fitness fixed during the 
adaptive walk  increases with the number of substitutions and with initial
fitness in all extreme value domains, the fitness difference 
(\ref{rateinfL}) between
successive steps depends on how fast the fitness distribution
(\ref{gpd}) decays (refer Figs. \ref{Fig1} and \ref{Fig2}). In the
Weibull domain, the fitness  
benefits decrease as the walk proceeds or the starting fitness is
increased. In contrast, in the
Fr{\'e}chet domain, increasing adaptive substitutions or initial fitness
leads to  increasing fitness gain. This
behavior of fitness increments is robust with respect to fitness
correlations as attested by the insets of Figs.~\ref{Fig1} and
\ref{Fig2}, and holds irrespective of whether the correct asymptotic
behavior of the fixation probability is taken into account. 
 To get some insight into the behavior of average fitness difference
  with initial fitness, we recall that the selection coefficient is
  bounded above for truncated distributions but not for the unbounded
  ones. As a result, in the Weibull domain, we have $f_1-f_0 \leq u
  -f_0$ which suggests that $\overline{\Delta f_1}$ decreases 
  with $f_0$. This is in contrast to the behavior in 
  the Fr{\'e}chet domain where the selection coefficient is large (and
  positive) and hence $f_1-f_0 \propto f_0$ which
  increases linearly with $f_0$. 
Treating the fitness at the $J$th step as the
initial fitness for the next step, the patterns during the course of
the walk can also be understood.  

We believe that experimental measurements of fitness 
difference as a function of the initial fitness in a population
evolving under strong selection-weak mutation conditions can give an insight
into the domain of the DBFE. Although negative 
correlation between initial fitness and fitness gain has been observed
\citep{Bull:2000,Maclean:2010} and increasing fitness gain in successive
steps has been seen in small populations \citep{Burch:1999}, how these
results correlate with the tail behavior of the beneficial mutations
in these studies is not known. On the other hand, the adaptive walk
properties have not been studied in experiments that measure the
exponent $\kappa$ 
\citep{Sanjuan:2004a,Rokyta:2005,Kassen:2006,Maclean:2010,Rokyta:2008,Bataillon:2011,Schenk:2012}. In \citet{Sousa:2012}, since the
local fitness optimum to which the population approaches is fixed, a truncated
fitness distribution is expected. It would be interesting to check if
our prediction that the fitness difference decreases with increasing
initial fitness for bounded distributions is supported by the fitness
data in their experiment.

We also studied 
the behavior of average selection coefficient and obtained a general
result that in all the three extreme value domains, it decreases
during the course of the walk and with initial fitness and fitness
correlations, see Fig.~\ref{Fig3} (also refer 
\citet{Kryazhimskiy:2009}). Since the fitnesses are closely related on
correlated 
fitness landscapes, the fitness difference and hence selection
coefficient is expected to decrease with increasing
correlations. The behavior with initial fitness can also be
rationalised using the fact that the fitness difference grows at most
linearly with initial fitness and therefore for large $f_0$, selection
coefficient decreases. These results are consistent with those
obtained in the experimental studies on {\it Aspergillus nidulans}
\citep{Schoustra:2009,Gifford:2011} in which the mean selection coefficient is
observed to decrease as the walk proceeds, and 
{\it Escherichia coli} \citep{Sousa:2012} in which  
the mean selective effect is found to be larger for poorer initial
condition.

\subsection*{Length of the adaptive walk}

On uncorrelated fitness landscapes, we find that the walk length
depends logarithmically on the initial fitness rank but an analytical
understanding 
of the trends in the walk length is presently missing. 
The effect of fitness 
correlations is to increase the walk length since correlated
fitness landscapes are less rugged
\citep{Orr:2006a,Jain:2011d,Filho:2012}. 
The adaptive walk length
has been seen to decrease with increasing initial fitness in some 
experiments \citep{Rokyta:2009,Sousa:2012} but contrary to the results
presented here, \citet{Gifford:2011} found 
the average walk length to be 
insensitive to the starting fitness. 
The latter result was rationalised by noting that the size of the selective 
effect at the first step increased with decreasing initial fitness
\citep{Gifford:2011}.  
However as discussed above, selection coefficient decreases with
increasing initial fitness and yet the walk length depends on the initial fitness.  
Thus the weak
dependence of walk length on starting fitness is not explained by
increased mutational size. 

Insensitivity of walk length to initial conditions and
decrease in fitness gain are expected for any $\kappa$ if, to
start with, the population is close to a local fitness optimum. 
In  \citet{Gifford:2011}, the  
distance to the fitness peak has been gauged by the ratio of initial
fitness to that of the known local fitness optimum in the experiment. However  
from (\ref{rank}) and (\ref{largestB}), we see that the initial rank,
which gives the number of better mutants available at the start of the
adaptation process \citep{Orr:2002}, is not linear in 
$f_0/{\tilde f}$ (except for $\kappa=1$) and depends strongly on the exponent
$\kappa$. For a sequence of length $L=10^3$
and an initial fitness half of that of a local fitness peak,
the initial rank is given by $500, 31.6, 1.4$ for $\kappa=-1, 0, 2$
respectively, which increases to $750, 178, 1.9$ when the initial
fitness is decreased to one quarter of the fitness of a local
maximum. 
Then in the absence of information about the 
exponent $\kappa$ in the experimental set up of \citet{Gifford:2011},
it is not clear if the population is sufficiently far from the fitness
optimum. 
Measurement of walk length for 
initial fitnesses smaller than those used in the experiment of
\citet{Gifford:2011} and of fitnesses fixed during evolution may help in 
understanding the properties of adaptation in this experiment. 

 \subsection*{Limitations of this work}

In the experiments discussed above, the 
  population size is $> 10^4$ \citep{Rokyta:2009,
    Schoustra:2009,Gifford:2011, Bataillon:2011, Sousa:2012}, the
  smallest selection coefficient detected is $\sim 10^{-3}$
  \citep{Gifford:2011,Sousa:2012} and the 
  mutation rate per base pair 
  for the 
  microbes used in the experiments namely, bacteriophage $\phi X174$ \citep{Rokyta:2005}, 
  \textit{Escherichia coli} \citep{Sousa:2012},  \textit{Aspergillus
    nidulans} \citep{Schoustra:2009}  is of the order $10^{-7}-10^{-11}$ \citep{Drake:1998}. 
Thus these experiments are in the strong selection-weak mutation
regime where adaptive walk model studied here is defined. However when the
  population size is large enough that the weak mutation condition
  fails, clonal
  interference occurs in which two or more independent beneficial mutations
  arise in the population and compete with each other for dominance
  \citep{Gerrish:1998,Desai:2007a,Barrick:2009,Gordo:2012}.  It would
  be interesting to check  
  whether the trends in the fitness difference discussed here are also
  exhibited by populations 
  with competing beneficial mutations especially during the early
  adaptation stage. However in the late
adaptation regime when the population has access to relatively few
beneficial mutations, we may expect the fitness difference trends
observed here to hold.

{\it Acknowledgements:} We thank J. Krug, S. Kryazhimskiy,  
C. J. Marx and S. N. Majumdar for useful discussions during various 
stages of this work. We also thank an anonymous reviewer for many helpful
comments and suggestions. 


\section*{Appendix :Fitness fixed and selection coefficient on uncorrelated fitness
  landscapes}
\label{AppA}

To find the average fitness and selection coefficient, we 
consider the probability distribution ${\cal P}_J(f|f_0)$ of the population fitness $f$
at the $J$th step of the adaptive walk, given that it started with 
fitness $f_0$. On uncorrelated fitness landscapes, it obeys the
following recursion equation \citep{Flyvbjerg:1992,Jain:2011d} 
\be
{\cal P}_{J+1}(f|f_0)= \int_{f_0}^{f} dh ~T(f \leftarrow h)~(1-q^L(h))~
{\cal P}_J(h|f_0)~,~J \geq 0
\label{main}
\ee
where $q(f)=\int_{0}^f dg~p(g)$ is the probability of having a
fitness less than $f$ and the initial condition ${\cal P}_0(h|f_0)=\delta(h-f_0)$ corresponds to a monomorphic population.  Equation (\ref{main}) 
simply means that the population moves from fitness $h$ to a higher
fitness $f$ at the next step with probability (4) provided at 
least one fitter  mutant is available, the probability of whose is given by 
$1-q^L(h)$.

The average fitness fixed at the $J$th step is given by ${\bar f}_J
(f_0)=\int_{f_0}^u df ~f ~{\cal 
  P}_{J}(f|f_0)$. Far from a local fitness peak, the average fitness
fixed for a sequence of length 
$L$ is well approximated by the corresponding quantity for an
infinitely long sequence \citep{Jain:2011d}. Then on using
(\ref{main}) in the limit $L \to \infty$ in the definition of the
average fitness ${\bar f}_{J+1}$, 
we have
\bea
{\bar f}_{J+1}(f_0) 
=  \int_{f_0}^u dh ~\Phi_J(h|f_0) ~ \int_h^u df~f~T(f \leftarrow h)
\label{fJinf}
\eea
where $ \Phi_J(f|f_0) \equiv \textrm{Lim}_{L \to \infty} {\cal
  P}_J(f|f_0)$, and we have interchanged the order of integration to
arrive at the  last equation. 
For a given $h$, the transition probability (4) varies as
$(f-h)~p(f)$ for $f \ll 3 h/2$ (small selection coefficient) and
$p(f)$ when selective effects are large. As the dominant 
contribution to the inner integral in (\ref{fJinf}) comes from the
large-$f$ behavior of the integrand, the integral over $f$ is seen to
be proportional to the mean of the fitness distribution $p(f)$ 
which, we recall, is undefined for $\kappa \geq 1$. 
This result means that the fitness fixed is independent of the sequence
length $L$ for $\kappa < 1$, but increases with $L$ otherwise. 
The average selection coefficient fixed at step $J$ also exhibits a
similar behavior. To see this, consider the distribution ${\cal
  S}_J(s|f_0)$ of selection coefficient 
$s$ at the $J$th step in the walk which can be determined using
(\ref{main}) for an infinitely long sequence as 
\bea
{\cal S}_J(s|f_0) &=& \int_{f_0}^u df \int_{f_0}^u dh~ \delta \left(s-
\frac{f-h}{h} \right) ~ T(f \leftarrow h)~ {\Phi}_{J-1}(h|f_0) \\
&=& \int_{f_0}^{\frac{u}{s+1}} dh  ~h ~ T( h (s+1) \leftarrow
h)~{\Phi}_{J-1}(h|f_0) ~,~ J \geq 1
\label{Sdefn}
\eea
In the last equation, the upper limit of the integral is obtained
using the fact that the fitness $f$ at the $J$th step can not exceed
$u$.  Then the average selection coefficient can be written as 
\bea
{{\bar s}}_J  (f_0) 
&=& \int_{f_0}^{u} dh ~h~{\Phi}_{J-1}(h|f_0) \int_0^{\frac{u}{h}-1} ds
~ s ~T(h (s+1) \leftarrow h)
\eea
Since the inner integral over $s$ in the last equation is undefined for
$\kappa \geq 1$ for the same reasons as described above for average
fitness, we find that the average selection coefficient also undergoes 
a transition at $\kappa=1$. 

 The fitness improvement $\overline{\Delta
  f_J}$ during the successive steps is defined as 
\be
\overline{\Delta f_J}={\overline {f_J- f_{J-1}}}
\label{finL}
\ee
where the overbar represents averaging over only those walks that
reach the $J$th step for a sequence of finite length. For infinitely
long sequences, as the 
$J$th step is definitely taken, we have $\overline{\Delta f_J}={\bar
  f}_J-{\bar f}_{J-1}$. Thus it is sufficient to study the behaviour
of the fitness fixed at each step which we discuss next. 

{\it Gumbel domain:} On performing the inner integral in (\ref{fJinf}) for $\kappa \to 0$, we get
\be
{\bar f}_{J+1}(f_0) 
=  \int_{f_0}^\infty dh ~\Phi_J(h|f_0) ~ \frac{h^2+4 h+2}{h+2}
\ee
The above equation does not close in the average fitness fixed i.e. the RHS contains the average of quantities which can not be written in terms of ${\bar f}_J$. However for large initial fitness $f_0$, as $h \gg 1$, we can write 
\bea
{\bar f}_{J+1}(f_0) 
&=&  \int_{f_0}^\infty dh ~\Phi_J(h|f_0) ~ \left(h+2 +
{\cal O}(h^{-1})\right)\\
&=& {\bar f}_{J}+2 +{\cal O}\left( \overline{f_J^{-1}} \right)
\label{fJapp}
\eea
where we have used that the adaptive walk 
goes on indefinitely for an 
infinitely long sequence \citep{Jain:2011d}.  
As the average fitness increases during adaptation, one may expect the
average of inverse fitness to decrease. Neglecting the last term on
the RHS of (\ref{fJapp}), we immediately find the solution of the
resulting recursion equation to be 
\bea
{\bar f}_J = 2 J +f_0 
\label{Gdiff}
\eea

{\it Weibull domain:} The inner integral in (\ref{fJinf}) can be done
exactly, but the resulting expression is too complicated and we omit
the general expression here. For the special case of $\kappa=-1$, we
get 
\be
{\bar f}_{J+1}(f_0) 
=  \int_{f_0}^1 dh ~\Phi_J(h|f_0) ~\frac{2 e^{2/h} (1- h)^2 + 
 e^2 h (2 + h) (\Gamma(2,2 - 2/h)-1)}{e^{2/h} (6 h-4)-2 e^2 h}
\ee
where $\Gamma(a,x)$ is the incomplete gamma function
\citep{Abramowitz:1964}. The above equation 
demonstrates that as in the Gumbel domain, the recursion relation for
${\bar f}_{J}$ does not close here also. We note from Fig.~Fig. 3 that the
selection coefficient is well below one when the 
initial fitness $f_0$ is close to $u$. In the inner integral on the
RHS of (\ref{fJinf}), the selection coefficient is smaller than half  
if the fitness $f \ll 3 h/2$ which is ensured if $f_0 > 2 u/3$. 
These observations suggest that in the Weibull domain, the small
selection coefficient can be assumed to be small. 
On using (6) in (\ref{fJinf}), we find that the
recursion equation closes in the average fitness and given by ${\bar
  f}_{J+1} =a_- {\bar f}_J+b_-$ where 
\bea
a_- &=& (1-2 \kappa)^{-1} \\
b_- &=& 2 (1-2 \kappa)^{-1}
\eea
On iterating the recursion equation, we find the average fitness to be 
\be
{\bar f}_J=a_-^J f_0 + \frac{b_-}{1-a_-}~(1-a_-^{J})
\label{Wdiff}
\ee
It is evident that for negative $\kappa$, the coefficient $a_- < 1$.  
It is easily verified that (\ref{Gdiff}) is obtained from the above
equation when $\kappa \to 0$.

{\it Fr{\'e}chet domain:} For $\kappa < 1$ and large $f_0$, proceeding
in manner similar to that in the Gumbel domain, we find that the 
average fitness at step $J$ is of the form ${\bar f}_{J+1} \approx a_+ {\bar
  f}_J+b_+$ 
where 
\bea
a_+ &=& \frac{\kappa -e^2 (1-\kappa)  E_{\frac{1}{\kappa }}(2)}{2 e^2 \kappa
  (1-\kappa)  E_{\frac{1}{\kappa }}(2)} \label{kpos_fj2}\\
b_+ &=& \frac{\kappa-e^2 (1+\kappa) E_{\frac{1}{\kappa}}(2)-2 e^4  \kappa (1-\kappa)E^2_{\frac{1}{\kappa}}(2)}{2 e^4 \kappa^2 (1-\kappa)  E^2_{\frac{1}{\kappa}}(2)}
\eea
and $E_n(x)$ is the exponential integral \citep{Abramowitz:1964}. For $\kappa \to 0$, using the large $n$
representation of $E_n(x)$ 
\citep{Abramowitz:1964} in the above expressions for $a_+$ and $b_+$, it
can be checked that the result 
(\ref{Gdiff}) in the Gumbel domain is obtained. 

For $\kappa\geq1$ where the mean of the fitness distribution becomes
infinite, we work with a sequence of 
finite length to find how the average fitness diverges with $L$. Since
the adaptation process is over when the fitness fixed is of the order
of the average fitness of a local fitness
optimum, we truncate the fitness distribution  (1)
at the average fitness of a local maximum  \citep{Sornette:2000}. For
uncorrelated 
fitness landscapes, on replacing the upper limit $u$ by ${\tilde f}$
in (\ref{fJinf}), we obtain 
\be
{\bar f}_{J+1}(f_0) 
=  \int_{f_0}^{\tilde f} dh ~\Phi_J(h|f_0) ~ \int_h^{{\tilde f}} df~f~T(f \leftarrow h)
\label{kgt1}
\ee
The inner integral in the above equation scales as 
${\tilde f}^{\frac{\kappa-1}{\kappa}}$ (or $L^{\kappa-1}$, due to
 (3)) for $\kappa > 1$ and $\ln L$ for
$\kappa=1$. Thus the  
fitness fixed at any step in the adaptive walk  depends
  strongly on the sequence length, when the mean of the 
fitness distribution is infinite.

To summarise, we find that the
final fitness $u$ is  approached exponentially for 
bounded distributions, but the fitness increases linearly with the number of 
substitutions for exponentially  
distributed fitnesses and exponentially for unbounded distributions 
with $0 < \kappa < 1$. For zero initial fitness, 
the results (\ref{Gdiff}) and (\ref{Wdiff}) for average fitness in the
Gumbel and Weibull domain respectively match Eq. 33 of  
\citet{Joyce:2008} for high initial rank and $\kappa \leq 0$, but in the Fr{\'e}chet domain, our
result differs from that of \citet{Joyce:2008} who work in the 
small selection coefficient approximation.

\newpage

\end{document}